\documentclass[aps,print,twocolumn,showpacs]{revtex4}
\usepackage{eurosym}
\usepackage{amsfonts}
\usepackage{amsmath}
\usepackage{amssymb}
\usepackage{graphicx}

\begin{document}

\title{Symmetry breaking of quantum droplets in a dual-core trap}
\author{Bin Liu$^{1}$, Hua-Feng Zhang$^{2}$, Rong-Xuan Zhong$^{1}$, Xi-Liang
Zhang$^{1}$, Xi-Zhou Qin$^{1}$, Chunqing Huang$^{1}$, and Yong-Yao Li$^{1}$$%
^,$$^{3}$}
\email{yongyaoli@gmail.com}
\author{Boris A. Malomed$^{3}$}
\affiliation{$^{1}$School of Physics and Optoelectronic Engineering, Foshan University,
Foshan 528000, China }
\affiliation{$^{2}$School of Physics and Optoelectronic Engineering, Yangtze University,
Jingzhou 434023, China}
\affiliation{$^{3}$Department of Physical Electronics, School of Electrical Engineering,
Faculty of Engineering, and Center for Light-Matter Interaction, Tel Aviv
University, Tel Aviv 69978, Israel}
\pacs{03.75.Lm, 05.45.Yv}

\begin{abstract}
We consider the dynamical model of a binary bosonic gas trapped in a
symmetric dual-core cigar-shaped potential. The setting is modeled by a
system of linearly-coupled one-dimensional Gross-Pitaevskii equations with
cubic self-repulsive terms and quadratic attractive ones,which represent the
Lee-Huang-Yang corrections to the mean-field theory in this geometry. The
main subject is spontaneous symmetry breaking (SSB) of \textit{quantum
droplets} (QDs), followed by restoration of the symmetry, with respect to
the identical parallel-coupled trapping cores, following the increase of the
QD's total norm. The SSB transition and inverse symmetry-restoring one form
a \textit{bifurcation loop}, whose shape in concave at small values of the
inter-core coupling constant, $\kappa $, and convex at larger $\kappa $. The
loop does not exist above a critical value of $\kappa $. At very large
values of the norm, QDs do not break their symmetry, featuring a flat-top
shape. Some results are obtained in an analytical form, including an exact
front solution connecting asymptotically constant zero and finite values of
the wave function. Collisions between moving QDs are considered too,
demonstrating a trend to merger into breathers.
\end{abstract}

\maketitle

\section{Introduction}

Recently, a new type of self-bound quantum liquid states, in the form of
three-dimensional (3D) droplets, was created experimentally in dipolar
bosonic gases of dysprosium \cite{Schmitt2016} and erbium \cite{Chomaz2016},
as well as in mixtures of two atomic states of $^{39}$K with contact
interactions \cite{Cabrera2018}, following the theoretical proposal
elaborated in Refs. \cite{Petrov2015} and \cite{Petrov2016}. These quantum
droplets (QDs) are formed by the balance of attractive forces, which drive
the collapse of the quantum gases in the mean-field approximation, and the
repulsive force induced by quantum fluctuations around the mean-field
states, which is represented by the quartic Lee-Huang-Yang (LHY) corrections
\cite{LHY} to the respective Gross-Pitaevskii equations (GPEs) with the
usual cubic terms. In the dysprosium and erbium gases, the attractive force
is generated by the dipole-dipole interactions, as it was analyzed in detail
\cite%
{Wachtler2016,Wachtler20162,Daillie2016,PRL119_050403,PRL116_215301,PRL120_160402,PRA97_011602,unstable-vort-DD,PRA94_033619}%
, while in the binary mixture it is provided by the inter-component
attraction, which can be made slightly stronger than the intra-component
repulsion by means of the Feshbach resonance, see further details in Refs.
\cite%
{Luca-Rabi,Staudinger2018,Cikojevic2018,Astrakharchik2018,PRA97_053623,PRA97_063616,Cabrera2,Inguscio}%
. Recently, the formation of QDs in Bose-Fermi mixture under the action of
the spin-orbit coupling (SOC)\ \cite{PRA98_023630}, and a possibility to
create similarly built photonic droplets \cite{PRA98_053835} have also been
predicted.

QDs are made of extremely dilute quantum fluids \cite{Tolra2016}. The
droplets may be considered as soliton-like objects, with the unique property
of \textit{stability} in 2D and 3D geometries, where usual nonlinear models
give rise to solitons that are subject to strong instabilities \cite%
{Review1,Review2} [an exception is provided by pairs of GPEs with SOC terms,
which predict absolutely stable 2D solitons, i.e., the system's ground
states \cite{HS1}, and metastable 3D ones \cite{HPu}]. Accordingly, stable
QDs offer potential use in various applications, such as matter-wave
interferometry \cite{interfero1,interfero2,interfero3} and manipulations of
quantum information \cite{info}. Furthermore, it was recently predicted that
2D QDs (whose effective nonlinearity is different from the above-mentioned
quartic form, amounting to cubic terms multiplied by a logarithmic factor
\cite{Petrov2016}) with embedded vorticity $S=1,2,3,...$ may be stable too,
up to $S=5$ \cite{2DvortexQD}. A related result is the stability of 2D QDs
of the \textit{mixed-mode} type (mixing vortical and zero-vorticity
constituents), formed by the SOC effect \cite{SOCQD}. Full 3D QDs with
embedded vorticity $S=1$ and $2$ have also been predicted to have stability
domains in the respective parameter space \cite{Kartashov}.

One of fundamental aspects of the soliton phenomenology is spontaneous
symmetry breaking (SSB) of self-trapped modes in symmetric two-component
systems. In particular, the SSB of optical solitons was considered in
various settings \cite%
{Wabnitz,Snyder1991,Trillo1998,Maimistov,Zhigang,Herring2007,Albuch2007,Zhijie}%
, including coupled lasers \cite{laser1,laser2} and metamaterials \cite{meta}
(see also a collection of articles on this topic \cite{volume}, and a review
in Ref. \cite{Tlidi}). Applications of this effect, such as design of
power-switch devices based on soliton light propagation in fibers, were
proposed \cite{Trillo1998,Tlidi}. In Bose-Einstein condensates (BECs), SSB
of matter-wave solitons has also been considered in many configurations \cite%
{BEC1,BEC2,Gubeskys2007,Matuszewski2007,Yongyao2013,Tlidi,Yongyao2017,Zhaopin2017}%
, but not, as yet, for QDs. In this work, we address effectively
one-dimensional QDs in the binary bosonic gas loaded in a symmetric
double-core cigar-shaped potential. Unlike the usual SSB mechanism for
matter-wave solitons, which is induced by mean-field interactions, the SSB
of QDs in this system is driven by the interplay of the mean-field and LHY\
terms.

The rest of the paper is structured as follows. The model is introduced in
Sec. II, where some analytical results are presented too, such as an exact
solution for a front interpolating between zero and an asymptotically
constant wave function. Basic numerical results for the SSB of QDs are
reported in Sec. III, which, in addition, includes some approximate
analytical results related to the numerical ones. Collisions of
two-component QDs are addressed in Sec. IV. The paper is concluded by Sec. V.

\section{The model}

\begin{figure}[h]
{\centering\vspace{-9.5cm} \includegraphics[width=2.3\columnwidth]{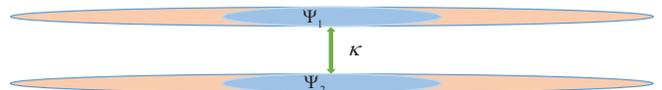}
\vspace{-0.5cm}}
\caption{(Color online) The schematic of the system: quantum droplets
(denoted by the blue color), which are formed in the Bose gas, are trapped
in the symmetric dual-core quasi-one-dimensional potential (the orange areas
remain empty). The parallel cores are coupled by hopping with rate $\protect%
\kappa $. }
\label{sketch}
\end{figure}

The system under the consideration is sketched in Fig. \ref{sketch}: QDs,
which are formed in the binary bosonic gas, are trapped in the nearly-1D
symmetric double-core potential, assuming, as usual, that the wave-function
components of the two species of the binary condensate are equal in each
core. Then, the system of linearly-coupled GPEs, including the LHY terms,
are written in the scaled form as \cite{Petrov2016,Astrakharchik2018}:%
\begin{align}
i\partial _{t}\Psi _{1}& =-\frac{1}{2}\partial _{xx}\Psi _{1}+g\left\vert
\Psi _{1}\right\vert ^{2}\Psi _{1}-\left\vert \Psi _{1}\right\vert \Psi
_{1}-\kappa \Psi _{2},  \notag \\
i\partial _{t}\Psi _{2}& =-\frac{1}{2}\partial _{xx}\Psi _{2}+g\left\vert
\Psi _{2}\right\vert ^{2}\Psi _{2}-\left\vert \Psi _{2}\right\vert \Psi
_{2}-\kappa \Psi _{1},  \label{Model}
\end{align}%
where $g\sim (g_{+-}+\sqrt{g_{++}g_{--}})/\sqrt{g_{++}g_{--}}>0$ is the
effective coefficient of the cubic repulsion \cite{Astrakharchik2018} ($%
g_{++,--}\ $and $g_{+,-}$\ are, respectively, strengths of the self- and
cross-interaction of the two components), and $\kappa >0$ is the hopping
rate which couples the parallel cores. By means of additional rescaling, we
fix $g\equiv 1$ in Eq. (\ref{Model}), making $\kappa $ the single control
parameter. The competition of the self-repulsive cubic and attractive
quadratic terms in Eq. (\ref{Model}) determines the formation of QDs in this
setting \cite{Astrakharchik2018}. Previously, a dual-core model with the
competition of cubic self-attraction and quintic repulsion in each core was
introduced in optics \cite{Albuch2007}.

A realistic model applicable to the experiment should include loss terms,
the main source of which are three-body collisions in the bosonic
condensate. In fact, the losses were analyzed in detail, in the present
contexts, in Refs. \cite{Cabrera2018} and \cite{Cabrera2} (including
supplemental materials of both publications). It was demonstrated,
theoretically and experimentally, that the losses, although they may be
conspicuous, allow one to work with solitons for quite a long time, which is
completely sufficient for the creation and observation of the QDs.

The total norm of the wave function, which is a dynamical invariant of the
model, being proportional to the total number of atoms in the dual-core
system, is
\begin{equation*}
N=N_{1}+N_{2}\equiv \int_{-\infty }^{+\infty }dx\left( |\Psi _{1}|^{2}+|\Psi
_{2}|^{2}\right) .
\end{equation*}%
Also conserved are the system's Hamiltonian and total momentum:%
\begin{gather}
H=\int_{-\infty }^{+\infty }dx\left[ \sum_{n=1,2}\left( \frac{1}{2}%
\left\vert \partial _{x}\left( \Psi _{n}\right) \right\vert ^{2}+\frac{1}{2}%
\left\vert \Psi _{n}\right\vert ^{4}-\frac{2}{3}\left\vert \Psi
_{n}\right\vert ^{3}\right) \right.  \notag \\
\left. -\kappa \left( \Psi _{1}\Psi _{2}^{\ast }+\mathrm{c.c.}\right) \right]
,  \label{H} \\
P=i\int_{-\infty }^{+\infty }dx\sum_{n=1,2}\Psi _{n}\partial _{x}\left( \Psi
_{n}^{\ast }\right) ,  \label{P}
\end{gather}%
where both $\ast $ and $\mathrm{c.c.}$ stand for the complex conjugation.

Stationary QDs with chemical potential $\mu $ are sought for as a solution
to Eq. (\ref{Model}) in the form of%
\begin{equation}
\{\Psi _{1},\Psi _{2}\}=\{\psi _{1},\psi _{2}\}e^{-i\mu t},
\label{Solutions}
\end{equation}%
with real stationary wave functions $\psi _{1}$ and $\psi _{2}$ obeying
equations (the prime stands for $d/dx$)
\begin{align}
\mu \psi _{1}& =-\frac{1}{2}\psi _{1}^{\prime \prime }+\psi _{1}^{3}-\psi
_{1}^{2}-\kappa \psi _{2},  \notag \\
\mu \psi _{2}& =-\frac{1}{2}\psi _{2}^{\prime \prime }+\psi _{2}^{3}-\psi
_{2}^{2}-\kappa \psi _{1}.  \label{Model2}
\end{align}%
Symmetric QD solutions of Eq. (\ref{Model2}), with $\psi _{1}=\psi _{2}$ and
the chemical potential taking values%
\begin{equation}
-2/9<\mu +\kappa <0,  \label{mu}
\end{equation}%
have the known form \cite{Astrakharchik2018}:%
\begin{gather}
\psi _{1,2}=\frac{-3\left( \mu +\kappa \right) }{1+\sqrt{1+(9/2)\left( \mu
+\kappa \right) }\cosh \left( \sqrt{-2\left( \mu +\kappa \right) }x\right) }
\notag \\
\equiv \psi _{\mathrm{symm}}(x).  \label{exact}
\end{gather}%
In the limit of $\left( \mu +\kappa \right) \rightarrow -0$, they take the
bell-shaped form,
\begin{equation}
\psi _{1,2}\approx \frac{-3\left( \mu +\kappa \right) }{2\cosh ^{2}\left(
\sqrt{-\left( \mu +\kappa \right) /2}x\right) }.  \label{small-mu}
\end{equation}%
In the opposite limit of
\begin{equation}
\mu +\kappa \rightarrow -2/9  \label{2/9}
\end{equation}%
[see Eq. (\ref{mu})], the soliton features an extended flat-top shape, with
a nearly constant intrinsic wave function,%
\begin{equation}
\psi _{1,2}\approx 2/3,  \label{2/3}
\end{equation}%
of size $\allowbreak $%
\begin{equation}
L\approx (3/2)\ln \left( \left( \mu +\kappa +9/2\right) ^{-1}\right) .
\label{log}
\end{equation}%
This flat-top wave form is bounded by two fronts, which are represented by
exact solutions of Eq. (\ref{Model2}), available precisely at $\mu +\kappa
=-2/9$:%
\begin{equation}
\psi _{1,2}=\frac{2/3}{1+\exp \left[ \pm \left( 2/3\right) \left(
x-x_{0}\right) \right] },  \label{front}
\end{equation}%
($x_{0}$ is an arbitrary shift of the coordinate), each interpolating
between $\psi _{1,2}=0$ and $\psi _{1,2}=2/3$, cf. Eq. (\ref{2/3}). The
energy of the front pattern, calculated as per Eq. (\ref{H}), is%
\begin{equation}
H_{\mathrm{front}}=8/81.  \label{Hfront}
\end{equation}%
A similar exact front solution of the GPE with the cubic-quintic
nonlinearity is known too \cite{Birnbaum}.

The SSB point is determined by the condition that the linearization of Eq. (%
\ref{Model2}) around the symmetric soliton produces a critical \textit{%
antisymmetric} eigenmode with the zero eigenvalue, $\delta \psi _{1,2}=\pm
\delta \psi _{0}$, which satisfies the linear equation \cite{Wabnitz,Tlidi},
\begin{equation}
\left( \mu -\kappa \right) \delta \psi _{0}=\left[ -\frac{1}{2}\frac{d^{2}}{%
dx^{2}}+3\psi _{\mathrm{symm}}^{2}(x)-2\psi _{\mathrm{symm}}(x)\right]
\delta \psi _{0}.  \label{linear}
\end{equation}%
In this work, we have obtained numerical asymmetric solutions of Eq. (\ref%
{Model2}) by means of the finite-difference method. Eq. (\ref{linear}) it is
used to predict the SSB point in an analytical approximation, see Eq. (\ref%
{approx}) below.

It is relevant to stress that all the modes which are antisymmetric or
asymmetric with respect to the parallel-coupled cores are spatial even ones,
i.e., $\psi _{1,2}\left( -x\right) =\psi _{1,2}(x)$. On the other hand, it
follows from Eq. (\ref{Model2}) that Eq. (\ref{linear}) with $\kappa =0$ has
an obvious exact solution, which, however, is spatially odd,
\begin{equation}
\delta \psi _{0}(x;\kappa =0)=\frac{\partial }{\partial x}\left[ \psi _{%
\mathrm{symm}}(x;\kappa =0)\right] .  \label{odd}
\end{equation}%
This fact implies that solutions asymmetric with respect to the two cores
cannot branch off from the symmetric ones at $\kappa =0$, keeping the
spatial parity.

The linear-stability analysis for the stationary states was performed by
adding small perturbations to solution (\ref{Solutions}):
\begin{align}
\Psi _{1}(x,t)& =\left[ \psi _{1}+\varepsilon w_{1}e^{iGt}+\varepsilon
v_{1}^{\ast }e^{-iG^{\ast }t}\right] e^{-i\mu t},  \notag \\
\Psi _{2}(x,t)& =\left[ \psi _{2}+\varepsilon w_{2}e^{iGt}+\varepsilon
v_{2}^{\ast }e^{-iG^{\ast }t}\right] e^{-i\mu t},  \label{perturbation}
\end{align}%
where $\varepsilon $ is a real infinitesimal amplitude of the perturbation
with eigenfunctions $w_{1}$, $w_{2}$, $v_{1}$ and $v_{2}$. The substitution
of expression (\ref{perturbation}) in Eq. (\ref{Model}) and subsequent
linearization leads to the eigenvalue problem in the matrix form,%
\begin{equation}
\left(
\begin{array}{cccc}
\hat{L}_{1} & -\kappa & \hat{L}_{3} & 0 \\
-\kappa & \hat{L}_{2} & 0 & \hat{L}_{4} \\
-\hat{L}_{3}^{\ast } & 0 & -\hat{L}_{1} & \kappa \\
0 & -\hat{L}_{4}^{\ast } & \kappa & -\hat{L}_{2}%
\end{array}%
\right) \left(
\begin{array}{c}
w_{1} \\
w_{2} \\
v_{1} \\
v_{2}%
\end{array}%
\right) =-G\left(
\begin{array}{c}
w_{1} \\
w_{2} \\
v_{1} \\
v_{2}%
\end{array}%
\right) ,  \label{Eig}
\end{equation}%
with operators%
\begin{gather}
\hat{L}_{1}=-\frac{1}{2}\partial _{xx}-\mu +2\left\vert \psi _{1}\right\vert
^{2}-\frac{3}{2}\left\vert \psi _{1}\right\vert ,  \notag \\
\hat{L}_{2}=-\frac{1}{2}\partial _{xx}-\mu +2\left\vert \psi _{2}\right\vert
^{2}-\frac{3}{2}\left\vert \psi _{2}\right\vert ,  \notag \\
\hat{L}_{3}=\psi _{1}^{2}-\frac{\psi _{1}^{2}}{2\left\vert \psi
_{1}\right\vert },  \notag \\
\hat{L}_{4}=\psi _{2}^{2}-\frac{\psi _{2}^{2}}{2\left\vert \psi
_{2}\right\vert }.  \label{L}
\end{gather}%
The linear eigenvalue problem based on Eq. (\ref{Eig}) can be solved by
means of the finite- difference method. As usual, the existence of an
imaginary part in a perturbation eigenfrequency, $G$, implies an instability.

\section{Symmetric and asymmetric quantum droplets}

\subsection{Generic numerical results}

\begin{figure*}[tph]
{\centering\includegraphics[width=1.8\columnwidth]{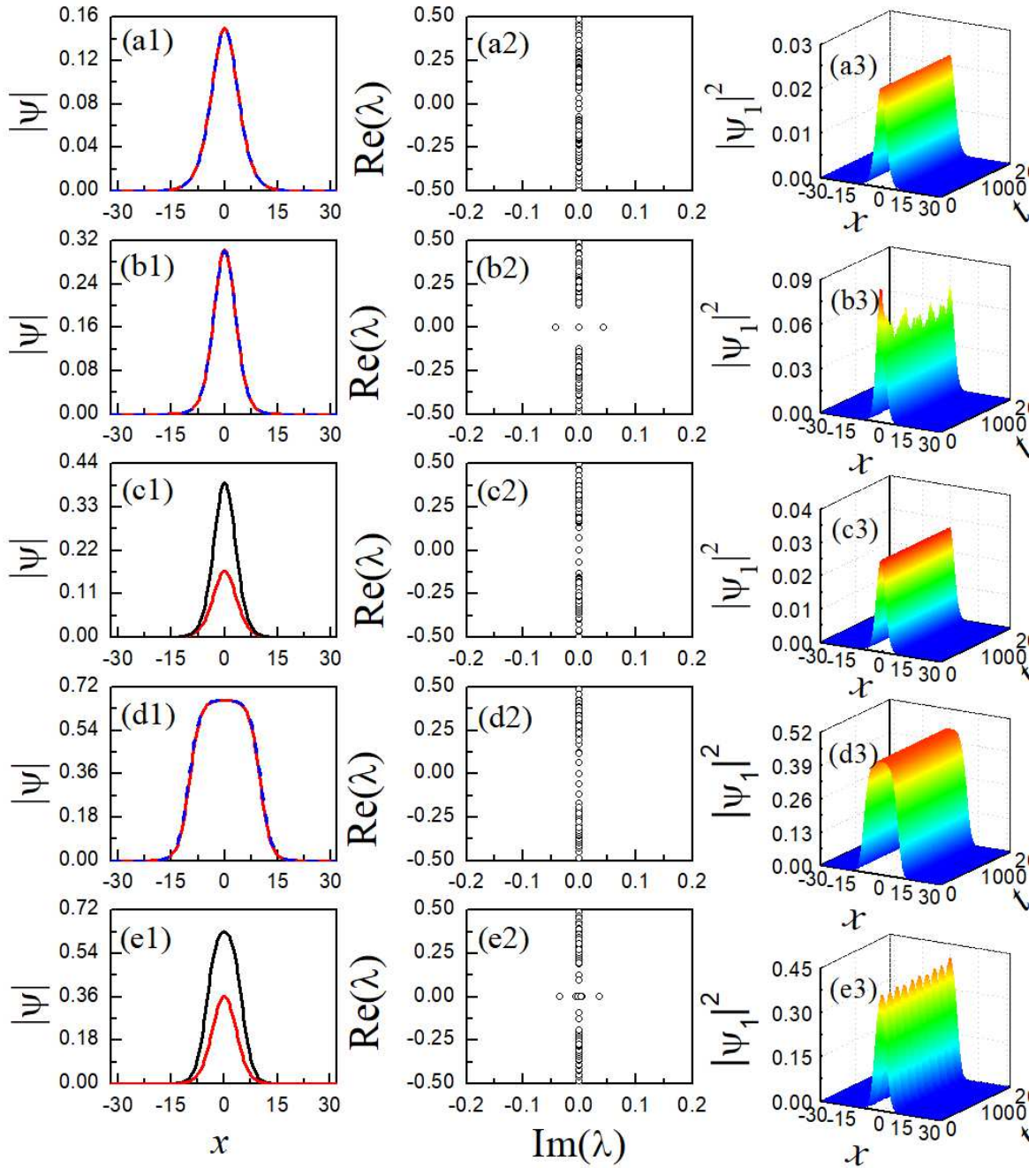}}
\caption{(Color online) Profiles of $\left\vert \protect\psi %
_{1}(x)\right\vert $ and $\left\vert \protect\psi _{2}(x)\right\vert $
components of the QDs, and the exact solution given by Eq. (\protect\ref%
{exact}), are shown by solid black , dotted red, and dashed gray curves,
respectively, for different values of the total norm, $N=0.3$, $1$, $1$, $20$%
, and $4$, severally, in panels (a1)-(e1). These examples of self-trapped
modes correspond to points a, b, c, and e, which are marked in Figs. \protect
\ref{fig1}(a,c) [point d is not marked, as the respective value of the norm,
$N=20$, is located beyond the frame of Fig. \protect\ref{fig1}(c)].
Perturbation eigenvalues for the corresponding symmetric [in (a1,b1,d1)] and
asymmetric [in (c1,e1)] QDs, and direct simulations of the perturbed
evolution of their $\Psi _{1}$ component, are displayed, respectively, in
panels (a2)-(e2) and (a3)-(e3). Parameters of Eq. (\protect\ref{Model}) are $%
\protect\kappa =0.05$ in panels (a3)-(d3) and $\protect\kappa =0.03$ in
panels (e3); the amplitude of small random perturbations in Eq. (\protect\ref%
{perturbation}) is $\protect\varepsilon =0.01$.}
\label{fig2}
\end{figure*}

Solutions for QDs which are symmetric and asymmetric with respect to the
coupled symmetric cores were produced with the help of the
imaginary-time-integration method \cite{Chiofalo2000,Jkyang2008}, applied to
Eq. (\ref{Model}). Figure \ref{fig2} displays typical examples of stable and
unstable QDs with different values of norm $N$. Similar to the situation in
the single-core model [see Eqs. (\ref{small-mu}) and (\ref{log})], QDs in
the present system feature spatial density profiles of two different types:
bell-shaped [see Figs. \ref{fig2}(a1)-(c1), (e1)] and flat-top ones [Figs. %
\ref{fig2}(d1)], for relatively small and large values of $N$, respectively.

Examples of stable symmetric and asymmetric QDs can be seen, respectively,
in Figs. \ref{fig2}(a1), (b1), (d1) and (c1), (e1). The asymmetry is
characterized by parameter%
\begin{equation}
\delta \equiv \left\vert \frac{N_{1}-N_{2}}{N_{1}+N_{2}}\right\vert .
\label{delta}
\end{equation}%
Accordingly, the SSB is characterized by dependences of $\delta $ on $N$ and
$\kappa $.

\begin{figure}[tph]
{\includegraphics[width=1\columnwidth]{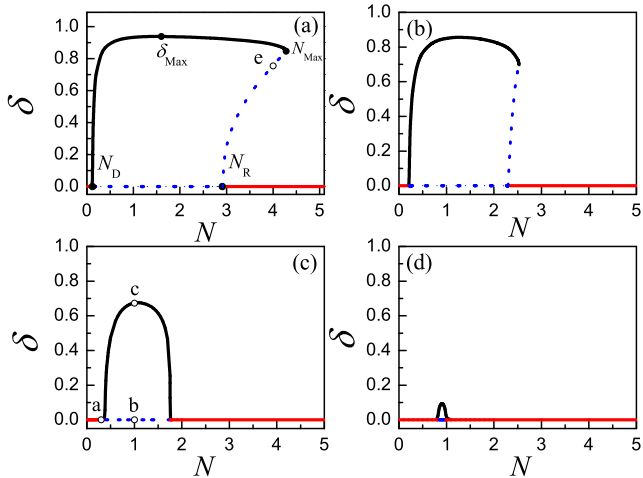}}
\caption{(Color online) A set of bifurcation diagrams for symmetric and
asymmetric QDs, in the plane of ($N$, $\protect\delta $), as found from
numerical solution of Eqs. (\protect\ref{Model2}) at different values of the
linear-coupling parameter, $\protect\kappa $: (a) $\protect\kappa =0.03$;
(b) $\protect\kappa =0.04$; (c) $\protect\kappa =0.05$; (d) $\protect\kappa %
=0.059$. Red, dotted blue, and black curves represent symmetric stable,
symmetric unstable, and asymmetric stable states, respectively.}
\label{fig1}
\end{figure}

First, in Fig. \ref{fig1} we produce $\delta (N)$ curves for different
values of $\kappa $. Due to the competition between the quadratic
self-attraction and cubic repulsion, they take the form of \textit{%
bifurcation loops }(somewhat similar to those in the cubic-quintic model
\cite{Albuch2007}), which exist at $\kappa \leq \kappa _{\max }\approx
0.0592 $. With the increase of $N$, the $\delta (N)$ curves first show the
SSB bifurcation of the supercritical (forward) type, driven by the quadratic
self-attraction, which is followed by a reverse symmetry-restoring
bifurcation, which occurs when the cubic repulsion becomes a dominant
nonlinear term. The latter bifurcation is of the subcritical (backward)
type, which lends the loop a concave shape, at $\kappa <\kappa _{0}\approx
0.044<\kappa _{\max }$. In the interval of $\kappa _{0}<\kappa <\kappa
_{\max }$, the symmetry-restoring bifurcation is supercritical, making the
loop a convex figure, which shrinks at $\kappa \rightarrow \kappa _{\max }$
and disappears at $\kappa =\kappa _{\max }$. It is relevant to mention that
the bifurcations of the subcritical and supercritical types are tantamount
to phase transitions of the the first and second kinds, respectively (see,
e.g., Ref. \cite{Zhijie} and references therein), thus predicting the
possibilities of these phase transition in the QDs trapped in the dual-core
potential.

The bifurcation loops are chiefly built of the QDs of the bell-shaped (%
\textrm{sech}) type, corresponding to relatively small and moderate values
of the norm, while the flat-top modes are found for large values of $N$, at
which the SSB is, in most cases, suppressed by the strong self-repulsive
nonlinearity.
\begin{figure}[tbp]
{\includegraphics[width=1\columnwidth]{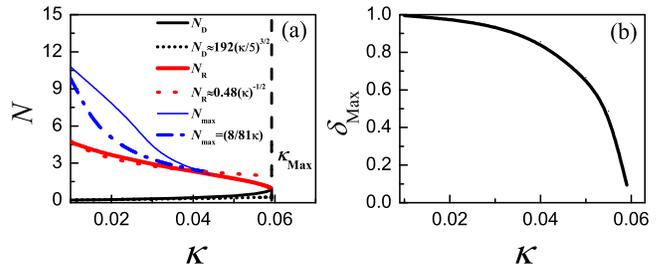}}
\caption{(Color online) (a) Values of the norm of the symmetric solution at
which the direct and reverse bifurcations occur ($N_{\mathrm{D}}$ and $N_{%
\mathrm{R}}$, respectively). The two curves merge and terminate at $\protect%
\kappa =\protect\kappa _{\max }\approx 0.0592$. The blue line shows the
largest value of the norm, $N_{\max }$, attained by asymmetric solitons, at
which the stable and unstable branches meet in the concave bifurcation loop,
see Figs. \protect\ref{fig1}(a,b). $N_{\max }$ merges with $N_{\mathrm{R}}$
at $\protect\kappa =\protect\kappa _{0}\approx 0.044$, the loops being
convex at $\protect\kappa >\protect\kappa _{0}$. The dotted black line,
short-dashed red line, and dashed-dotted blue line show analytical
approximation (\protect\ref{approx}), (\protect\ref{NR}) and (\protect\ref%
{Nmax}) for $N_{\mathrm{D}}$, $N_{\mathrm{R}}$, and $N_{\max }$,
respectively (coefficient $0.48$ in the analytical expression for $N_{%
\mathrm{R}}$ is a fitting constant). (b) The maximum values of the asymmetry
parameter, $\protect\delta $, of the solutions generated by the bifurcation
[see Eq. (\protect\ref{delta})], versus the linear-coupling parameter $%
\protect\kappa $.}
\label{fig3}
\end{figure}
\begin{figure*}[tph]
{\includegraphics[width=1.6\columnwidth]{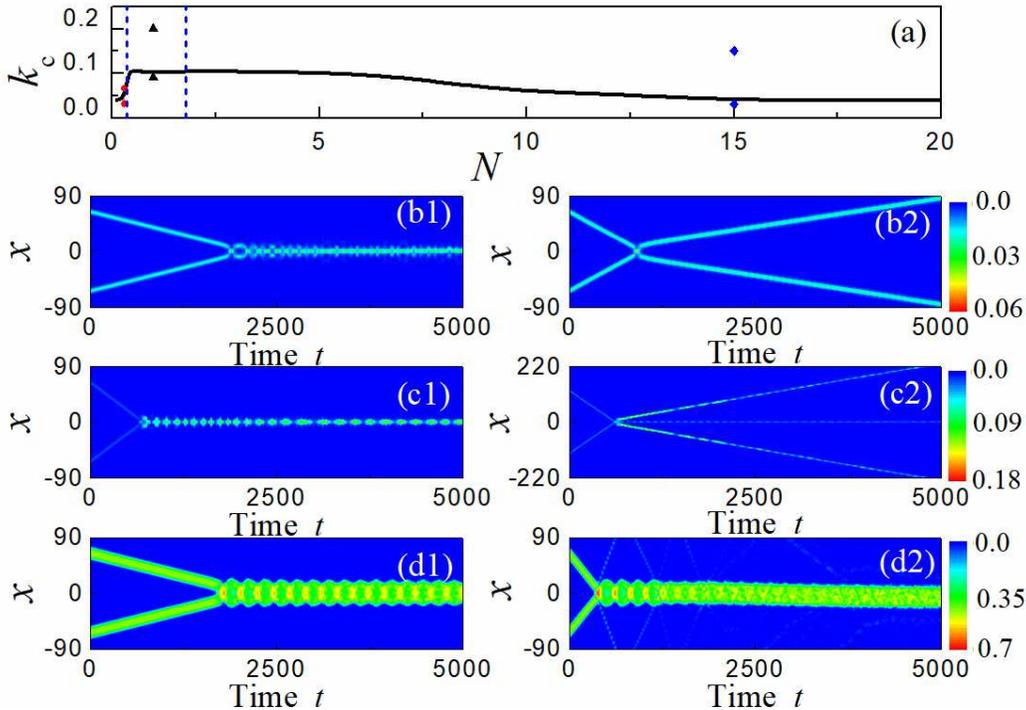}}
\caption{(Color online) (a) The value of the kick, $k_{c}$, which is a
boundary between the merger and passage of colliding QDs, launched as per
Eq. (\protect\ref{initial1}) with $D=64$, versus the total norm, $N$.
Typical examples of the density plots of the colliding droplets: (b1) $N=0.3$%
, $k=0.03$; (b2) $N=0.3$, $k=0.065$; (c1) $N=1$, $k=0.09$; (c2) $N=1$, $%
k=0.2 $; (d1) $N=15$, $k=0.03$; (d2) $N=15$, $k=0.15$. In this figure, $%
\protect\kappa =0.05$ is fixed. The colliding QDs are symmetric in (b1,b2)
and asymmetric in (c1,c2).}
\label{Collisions}
\end{figure*}

Results pertaining to the bifurcation loops are collected in Fig. \ref{fig3}%
. Namely, Fig. \ref{fig3}(a) displays values $N_{\mathrm{D}}$ and $N_{%
\mathrm{R}}$ of the total norm at the direct- and reverse-bifurcation
points, which merge at $\kappa =\kappa _{\max }$, and Fig. \ref{fig3}(b)
shows the largest value of asymmetry (\ref{delta}), $\delta _{\max }(\kappa
) $, as a function of the coupling constant. Figure \ref{fig3}(a) also
includes a plot (the blue line) showing the largest value $N_{\max }$ of $N$
attained in the concave loops, in the case of $\kappa <\kappa _{0}$.
Obviously, $N_{\max }\equiv N_{\mathrm{R}}$ at $\kappa _{0}<\kappa <\kappa
_{\max }$.

\subsection{Analytical results for the weakly-coupled system}

Approximate analytical results can be obtained in the limit of small $\kappa
$ and, accordingly, small $N_{\mathrm{D}}$. In this case, Eq. (\ref{linear})
with approximation (\ref{small-mu}) adopted for $\psi _{\mathrm{symm}}$, can
be solved exactly, using the well-known result from quantum mechanics,
similar to how this was done, in another context (dual-core optical fibers),
in Refs. \cite{Wabnitz,Tlidi}:%
\begin{equation}
\mu \approx -(13/5)\kappa ,~N_{\mathrm{D}}\approx 192\left( \kappa /5\right)
^{3/2}.  \label{approx}
\end{equation}%
In Fig. \ref{fig3}(a), the approximate dependence given by the second
equality in Eq. (\ref{approx}), is plotted by the dotted black line, showing
that it fits well to the numerically found dependence at $\kappa \lesssim
0.04$.

Further, both $N_{\max }$ and $N_{\mathrm{R}}$ diverge in the limit of $%
\kappa \rightarrow 0$, when one component in the asymmetric state (e.g., $%
\psi _{2}$) is vanishing, its amplitude in the flat-top states (which
correspond to large $N$) being%
\begin{equation}
\psi _{2}\approx 3\kappa ,  \label{small}
\end{equation}%
as it follows from Eqs. (\ref{Model2}), (\ref{2/9}), and (\ref{2/3}), while
\ a correction to the amplitude of the larger component is determined by the
conservation of the total norm:%
\begin{equation}
\psi _{1}\approx 2/3-(27/4)\kappa ^{2}.  \label{large}
\end{equation}

At $\kappa \rightarrow 0$,\ the value of $N_{\max }$ can be estimated,
taking into regard that the energy of the flat-top symmetric soliton is
larger than its single-component counterpart, with the same total norm, by
the amount equal to the front's energy (\ref{Hfront}), as the
single-component state includes only two fronts, unlike four ones in the
two-component state, and the energy (\ref{Hfront}) actually pertains to the
double front in the two-component symmetric state. Effectively, this is an
energy barrier which maintains the existence of the asymmetric soliton. On
the other hand, the weak linear coupling between the components in the
flat-top symmetric state of length $L$ corresponds to the negative energy,
which is%
\begin{equation}
H_{\mathrm{coupling}}\approx -(8/9)\kappa L,  \label{Hc}
\end{equation}%
according to Eq. (\ref{H}). The asymmetric\ state ceases to exist, by a jump
[like in Fig. \ref{fig1}(a)] under condition $H_{\mathrm{front}}+H_{\mathrm{%
coupling}}<0$, i.e., at $L>L_{\max }\approx \left( 9\kappa \right) ^{-1}$.
Eventually, the respective prediction for the largest norm, up to which the
asymmetric states exist at $\kappa \rightarrow 0$, is%
\begin{equation}
N_{\max }\approx 2(2/3)^{2}L_{\max }=8/\left( 81\kappa \right) .
\label{Nmax}
\end{equation}

Further, in the same limit of $\kappa \rightarrow 0$, value $N_{\mathrm{R}}$
at the reverse-bifurcation point also diverges, because, as shown above [see
Eq. (\ref{odd})], the SSB cannot take place in the form of an asymmetric
branch stemming from a symmetric one at $\kappa =0$ and some finite value of
$N$. A rough estimate for the divergence can be derived by noting that large
size $L$ of the symmetric QD is associated with a shift of the eigenvalue in
Eq. (\ref{linear}), $\kappa \sim L^{-2}$, hence the respective norm is
estimated as%
\begin{equation}
N_{\mathrm{R}}\approx 2(3/2)^{2}L\sim \kappa ^{-1/2},  \label{NR}
\end{equation}%
cf. Eq. (\ref{Nmax}).

The dotted black line, dash-dotted blue line, and short-dashed red line show
the analytical approximations given by Eqs. (\ref{approx}), (\ref{Nmax}) and
(\ref{NR}) for $N_{\mathrm{D}}$, $N_{\max }$, and $N_{\mathrm{R}}$,
respectively (the curve representing $N_{\mathrm{R}}$ is drawn with a
fitting parameter). It is seen that the analytically predicted values $N_{%
\mathrm{D}}$ and $N_{\mathrm{R}}$ agree well with their numerical
counterparts. The prediction given by Eq. (\ref{Nmax}) is less accurate, in
comparison with the numerical findings, but, still, it is qualitatively
correct.

\begin{figure}[tbp]
{\includegraphics[width=1\columnwidth]{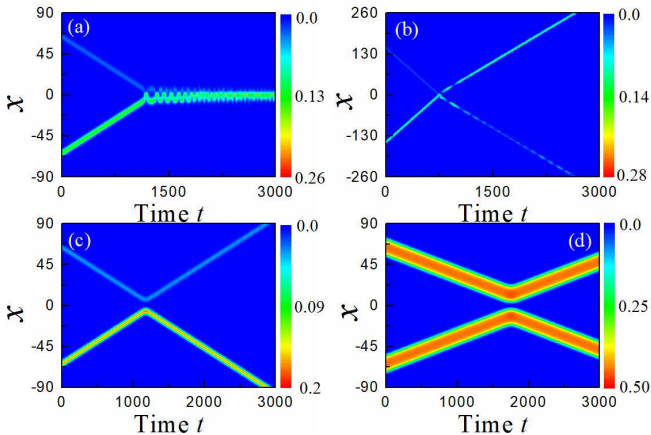}}
\caption{(Color online) Typical examples of density plots, in component $%
\Psi _{1}$, for collisions between two asymmetric QDs with opposite
placement of the larger and smaller components with respect to the two
cores. Panels (a) and (b) display, severally, strongly inelastic and
quasi-elastic collisions between in-phase solitons with norms $N=1$, the
respective kicks being $k_{\mathrm{a}}=0.05$ and $k_{\mathrm{b}}=0.20$.
Panels (c) and (d) show completely elastic collisions between out-of-phase
QDs, i.e., ones with $\protect\varphi =\protect\pi $ in Eq. (\protect\ref%
{initial1}), for $N=1$, $k=0.05$ and $N=15$, $k=0.03$, respectively. In this
figure, $\protect\kappa =0.05$ is fixed. }
\label{fig5}
\end{figure}

\section{Collisions of two-component droplets}

Once stable solitons are available in the Galilean-invariant system (\ref%
{Model}), it is relevant to explore collisions between them. In the
framework of the usual dual-core system with the cubic self-attractive
nonlinearity, collisions were studied in Ref. \cite{GangDing},
demonstrating, chiefly, merger of the colliding solitons into strongly
oscillating breathers, unless the solitons would pass through each other if
the collision velocity was very large.

We simulated the collisions, solving Eq. (\ref{Model}) with initial
conditions
\begin{equation}
\Psi _{1,2}(x,t=0)=\psi _{1,2}(x+D)e^{ikx}+\psi _{1,2}(x-D)e^{-ikx+\varphi },
\label{initial1}
\end{equation}%
where $\psi _{1,2}$ represent the stationary shape of two-component QDs, $%
\pm k$ is a kick, which sets two initial droplets, separated by distance $2D$%
, in motion with velocities also equal to $\pm k$, and $\varphi $ is the
initial phase difference between them.

The simulations demonstrate a trend to inelastic outcomes of the collisions
between the solitons in the in-phase configuration, i.e., with $\varphi =0$
in Eq. (\ref{initial1}). QDs of the bell (\textrm{sech})-shaped type merge
at relatively small values of $k$, and collide quasi-elastically (passing
through each other) at large $k$. A boundary value, $k_{c}$, which separates
the inelastic and elastic collisions is displayed in Fig. \ref{Collisions}%
(a), as a function of $N$, for $\kappa =0.05$. In particular, it
demonstrates that $k_{c}$ is smaller for symmetric bell-shaped QDs than for
asymmetric ones of the same type. This difference is explained by the fact
that the nonlinear interaction between larger components in the asymmetric
state is stronger than in the symmetric one, hence larger kinetic energy is
necessary to overcome the interaction and let the colliding QDs pass through
each other.

Figures \ref{Collisions}(b1,c1) and (b2,c2) show typical collision pictures
for $k<k_{c}$ and $k>k_{c}$, respectively. The examples displayed in panels
(b1,b2) and (c1,c2) correspond, respectively, to the red dot and black
triangle marks in panel (a). These pictures demonstrate that, when the
bell-shaped QDs pass through each other at $k>k_{c}$, the collisions
essentially perturb them. In particular, symmetric QDs emerge from the
collision with excited intrinsic oscillations and velocities different from
the original ones. In addition, colliding asymmetric bell-shaped QDs
generate an extra oscillating localized pulse (breather) with zero velocity.

We have also considered cross-symmetric collisions between two asymmetric
QDs, i.e., with opposite placements of the larger and smaller components
with respect to the two cores, as shown in Fig. \ref{fig5}(a)-(b). In this
case, strongly inelastic, quasi-elastic, and completely elastic outcomes are
observed too.

As the bell-shaped QDs carry over into flat-top ones with the increase of $N$%
, the newly generated quiescent breather grows larger, and eventually
absorbs almost all the norm of the colliding QDs, see an example in Fig. \ref%
{Collisions}(d2) (a similar outcome of collisions of single-component QDs
was reported in Ref. \cite{Astrakharchik2018}). Actually, this is a
different mechanism of the merger of colliding QDs, cf. panels (d1) and (d2)
in Fig. \ref{Collisions}, which correspond to the blue rhombic marks in Fig. %
\ref{Collisions}(a). We find that about $91\%$ of the total norm is absorbed
by the quiescent breather in Fig. \ref{Collisions}(d2).

Lastly, also similar to the results reported in Ref. \cite{Astrakharchik2018}
for the single-component model, completely elastic collisions (rebounds)
occur between the two-component QDs with opposite signs, i.e., $\varphi =\pi
$ in Eq. (\ref{initial1}), as shown in \ref{fig5}(c,d), and is observed in
other cases too.

\section{Conclusion}

The objective of this work is to study the SSB (spontaneous symmetry
breaking) of effectively one-dimensional QDs (quantum droplets) created in
the binary bosonic gas loaded in the dual-core trapping potential. The
matter-wave dynamics in this system is governed by the linearly-coupled GPEs
(Gross-Pitaevskii equations) with the cubic repulsive and quadratic
attractive nonlinear terms, the latter ones being represented by the LHY
(Lee-Huang-Yang) correction to the mean-field approximation. QDs in this
system feature bell (\textrm{sech})-shaped density profiles for smaller
values of total norm, $N$, and flat-top profiles for larger $N$. The SSB
bifurcation takes place with the increase of $N$, while the QDs keep the\
\textrm{sech }shape. Further increase of $N$ leads to the restoration of the
symmetry via the reverse bifurcation, hence the flat-top QDs, which realize
large values of $N$, are symmetric, in most cases. The resulting
bifurcations loops are concave and convex in the cases of small and larger
values of the inter-core coupling constant, and vanish when it exceeds a
critical value. Some results have been obtained in the analytical form -- in
particular, the exact solution was produced for a front separating zero and
finite constant values of the wave function, in the flat-top states.

Collisions between two-component QDs have been considered too. Unless the
colliding in-phase QDs move very fast, they tend to demonstrate inelastic
interactions, leading to their merger into breathers.

An interesting extension of the present analysis is to perform it for a
two-dimensional dual-core system, where the effective nonlinear terms in the
GPE is different, $\sim \left\vert \Psi \right\vert ^{2}\Psi \ln \left(
\left\vert \Psi \right\vert ^{2}\right) $ \cite{Petrov2016,2DvortexQD,SOCQD}%
. In that case, it will be possible to study the SSB not only in fundamental
two-component QDs, but also in ones with embedded vorticity, cf. Refs. \cite%
{Gubeskys2007,Nir}.

\begin{acknowledgments}
This work was supported by NNSFC (China) through grants No. 11874112,
11575063, 11547007, and by the Israel Science Foundation through grant No.
1287/17.
\end{acknowledgments}

\end{document}